\documentclass[twocolumn]{jpsj3}
\usepackage{txfonts}

\usepackage{graphicx}
\usepackage{dcolumn}
\usepackage{bm}
\usepackage{braket}
\usepackage{color}
\usepackage{accents}
\usepackage{mathrsfs}
\usepackage{bm}
\usepackage{here}
\usepackage{longtable}

\title{Quasiparticle on Bogoliubov Fermi surface and odd-frequency Cooper pair}

\author{Dakyeong Kim$^{1}$, Shingo Kobayashi$^{2}$, and Yasuhiro Asano$^{1,3}$}
\inst{
$^{1}$Department of Applied Physics, Hokkaido University, Sapporo 060-8628, Japan\\
$^{2}$RIKEN Center for Emergent Matter Science, Wako, Saitama 351-0198, Japan\\
$^{3}$Center of Topological Science and Technology,
Hokkaido University, Sapporo 060-8628, Japan
} 

\abst{
We discuss a close relationship between a quasiparticle on the Bogoliubov Fermi surface
 and an odd-frequency Cooper pair in a superconductor 
in which a Cooper pair consisting of two $j=3/2$ electrons forms the pseudospin-quintet even-parity
pair potential with breaking time-reversal symmetry.
It has been established in a single-band superconductor that 
a low-energy quasiparticle below the superconducting gap accompanies an odd-frequency Cooper pair.
In this paper, we show that an odd-frequency pair characterized by chirality coexists with a quasiparticle 
on the Bogoliubov Fermi surface. 
The symmetry of odd-frequency Cooper pairs is analyzed in detail by taking realistic 
pair potentials into account in a cubic superconductor. 
}

\begin{document}
\maketitle

\section{Introduction}
 Although gapped energy spectra at the Fermi level (zero energy) are a fundamental 
property of a superconductor, various quasiparticle states exist in subgap energy region
such as a quasiparticle state at nodes of an unconventional superconductor, that at a vortex core\cite{caroli:physlett1964,volovik:jetplett1993}, 
and a zero-energy state at a surface of a topologically 
nontrivial superconductor\cite{buchholtz:prb1981,hara:ptp1986}.
The last one has been a hot issue in condensed matter physics in this decade.
The Bogoliubov Fermi surface (BFS) represents a novel type of 
subgap state in an even-parity superconductor which breaks 
time-reversal symmetry spontaneously.\cite{agterberg:prl2017,brydon:prb2018}
The properties of a quasiparticle on BFS have attracted much attention because
the evidence for time-reversal symmetry breaking superconducting state has been reported in 
a number of unconventional superconductors such as
 uranium compounds\cite{heffner:prl1990,schemm:prb2015}, a ruthernate Sr$_2$RuO$_4$\cite{luke:nature1998}, 
a skutterudite PrOs$_4$Sb$_{12}$ \cite{aoki:prl2003}, and a transition metal dichalcogenide\cite{ganesh:prl2014}.
 The internal degree of freedom of an electron (orbital or sublattice) plays a key role
in stabilizing the BFS. 
Therefore the issue is related to multiband superconductivity in pnictides\cite{pnictide:hosono2008}, 
and in topological superconductivity\cite{hor:prl2010}, and superconductivity due to a Cooper pair 
consisting of two $j=3/2$ electrons\cite{brydon:prl2016,kim:sciadv2018}.
At present, however, the physics of a quasiparticle on the BFS has not been explored yet.

In a single-band unconventional superconductor, 
a topologically protected quasiparticle at zero energy always accompanies an odd-frequency 
Cooper pair\cite{berezinskii:jetplett1974,tanaka:prl2007,tanaka:jpsj2012} and modifies 
drastically low energy transport properties 
in superconducting proximity structures~\cite{tanaka:prb2004,asano:prl2006}. 
In the mean-field theory of superconductivity, a quasiparticle and a Cooper pair
are described by the normal Green's function and the anomalous Green's function, respectively.
Since the two Green's functions are related to each other through
the Gor'kov equation,  
the singularity in the normal Green's function at zero energy is compensated 
by an odd-frequency Cooper pairing correlation in the anomalous Green's function~\cite{schopohl:arxiv1998}.
In a two-band/orbital superconductor, the existence of the BFS enhances 
the subgap spectra in the density of states.
Therefore, a quasiparticle on the BFS is considered to generate an odd-frequency Cooper pair.
In this paper, we will show that this reasoning is correct.

 Following a general argument in Ref.~\citen{brydon:prb2018}, we consider a 
superconducting state preserving inversion symmetry 
and breaking time-reversal symmetry in a metal consisting of $j=3/2$ electrons near the Fermi level.
We assume that the pair potential belongs to $s$-wave pseudospin-quintet symmetry class.  
The stability of BFS is described by the negative values of the Pfaffian defined in terms of 
Bogoliubov-de Gennes (BdG) Hamiltonian\cite{agterberg:prl2017}. 
The appearance of an odd-frequency Cooper pair is described by the anomalous Green's function.
Odd-frequency pairing is a broader concept than the formation of the Bogoliubov Fermi surface.
Indeed, the latter requires both the time-reversal symmetry breaking superconducting states 
and the internal degree of an electron, whereas the former does not. 
Thus the Bogoliubov Fermi surface could be characterized by the appearance of an odd-frequency Cooper 
pair with special properties.
We analyze characters of odd-frequency Cooper pairs which appear 
under the promising pair potentials in a cubic superconductor.
An induced odd-frequency pair can be described by using chirality which is a measure that
represents the breaking down of time-reversal symmetry.

\section{Bogoliubov Fermi Surface }

 We begin our analysis with the normal state Hamiltonian adopted 
in Ref.~\citen{agterberg:prl2017}.
The electronic states have four degrees of freedom consisting of two orbitals of equal parity and spin 1/2. 
The Hamiltonian describes effective $j=3/2$ electrons
in the presence of strong spin-orbit interactions~\cite{luttinger:pr1955}
\begin{align}
\mathcal{H}_{\mathrm{N}} =& \sum_{\boldsymbol{k}} 
\Psi_{\boldsymbol{k}}^\dagger \, H_{\mathrm{N}}(\boldsymbol{k}) \, \Psi_{\boldsymbol{k}},\\ 
\Psi_{\boldsymbol{k}} =& \left[ c_{\boldsymbol{k}, 3/2}, 
c_{\boldsymbol{k}, 1/2}, c_{\boldsymbol{k}, -1/2}, c_{\boldsymbol{k}, -3/2}\right]^{\mathrm{T}},
\end{align}
where $\mathrm{T}$ means the transpose of a matrix and 
$c_{\boldsymbol{k}, j_z}$ is the annihilation operator of an electron at $\boldsymbol{k}$ with 
the $z$-component of angular momentum being $j_z$.
The normal state Hamiltonian is represented by
\begin{align}
H_{\mathrm{N}}(\boldsymbol{k}) =& \alpha \boldsymbol{k}^2 + \beta 
\left(\boldsymbol{k}\cdot \boldsymbol{J}\right)^2 - \mu=
 \xi_{\boldsymbol{k}} \, 1_{4\times 4}
+ \vec{\epsilon}_{\boldsymbol{k}} \cdot \vec{\gamma}
 \label{eq:hn_cubic}
\end{align}
with $\xi_{\boldsymbol{k}}= \epsilon_{\boldsymbol{k},0} -\mu$ and
\begin{align} 
\epsilon_{\boldsymbol{k}, 0} = &\left(\alpha+ \frac{5}{4}\beta\right) \boldsymbol{k}^2, \quad
\epsilon_{\boldsymbol{k},1} = \beta\, \sqrt{3}\,  k_x\, k_y, \\  
\epsilon_{\boldsymbol{k},2} =& \beta\, \sqrt{3}\,  k_y\, k_z, \quad  
\epsilon_{\boldsymbol{k},3} = \beta\, \sqrt{3}\,  k_z\, k_x, \\ 
\epsilon_{\boldsymbol{k},4} = &\beta\, \frac{\sqrt{3}}{2}\,  (k_x^2- k_y^2), \; \epsilon_{\boldsymbol{k},5} =  \frac{\beta}{2}\,  (2k_z^2 - k_x^2- k_y^2),
\end{align}
where $\vec{\epsilon}_{\boldsymbol{k}}$ represents a five-component vector and 
$\boldsymbol{J}=(J_x, J_y, J_z)$ is the spinor
with $j=3/2$, 
\begin{align}
J_x =& \frac{1}{2}\left[\begin{array}{cccc}
0 & \sqrt{3} & 0 & 0 \\
\sqrt{3} & 0 & 2 & 0\\
0 & 2 & 0 & \sqrt{3} \\
0 & 0 & \sqrt{3} & 0
\end{array}\right], \\
J_y =& \frac{1}{2}\left[\begin{array}{cccc}
0 & -i\sqrt{3} & 0 & 0 \\
i\sqrt{3} & 0 & -2i & 0\\
0 & 2i & 0 & -i\sqrt{3} \\
0 & 0 & i\sqrt{3} & 0
\end{array}\right], \\
J_z =& \frac{1}{2}\left[\begin{array}{cccc}
3 & 0 & 0 & 0 \\
0 & 1 & 0 & 0\\
0 & 0 & -1 & 0 \\
0 & 0 & 0 & -3
\end{array}\right].
\end{align}
The $4 \times 4$ matrices in pseudospin space are defined as
\begin{align}
\gamma^1=& \frac{1}{\sqrt{3}} (J_x J_y +J_y J_x), \quad
\gamma^2= \frac{1}{\sqrt{3}} (J_y J_z +J_z J_y), \\
\gamma^3=& \frac{1}{\sqrt{3}} (J_z J_x +J_x J_z), \quad
\gamma^4= \frac{1}{\sqrt{3}} (J_x^2- J_y^2), \\
\gamma^5=& \frac{1}{3} (2J_z^2 - J_x^2 -J_y^2),
\end{align}
and $1_{4\times 4}$ is the identity matrix.
They satisfy the following relations
\begin{align}
&\gamma^\nu\, \gamma^\lambda + \gamma^\lambda\, \gamma^\nu =2 \times  1_{4\times 4} \delta_{\nu, \lambda}, 
\label{eq:anti_c}\\
&\gamma^1\, \gamma^2\, \gamma^3\, \gamma^4\, \gamma^5=-1_{4\times 4},\\ 
&\{\gamma^\nu\}^\ast = \{\gamma^\nu\}^{\mathrm{T}}= U_T\, \gamma^\nu \, U_T^{-1},\quad U_T=\gamma^1\, \gamma^2,
\label{eq:ut}
\end{align}
where $U_T$ is the unitary part of the 
time-reversal operation $\mathcal{T}=U_T \, \mathcal{K}$ with $\mathcal{K}$ meaning complex conjugation.
The relations displayed in Eqs.~(\ref{eq:anti_c})-(\ref{eq:ut}) are particularly important to reach the conclusions.
The superconducting pair potential is represented as
\begin{align}
\Delta(\boldsymbol{k}) = \vec{\eta}_{\boldsymbol{k}} \cdot \vec{\gamma} \, U_T,
\end{align}
where a five-component vector $\vec{\eta}_{\boldsymbol{k}}$ represents
an even-parity pseudospin-quintet state. 
As a result of the Fermi-Dirac statistics of electrons, the pair potential is antisymmetric 
under the permutation of two electrons, (i.e.,  $\Delta^{\mathrm{T}}(\boldsymbol{k}) = - \Delta(\boldsymbol{-k}) $). 

The existence of BFS is discussed by using the Pfaffian $P(\boldsymbol{k})$ of the Bogoliubov-de Gennes 
Hamiltonian,
\begin{align}
H_{\mathrm{BdG}}(\boldsymbol{k})= \left[\begin{array}{cc}
H_{\mathrm{N}}(\boldsymbol{k}) & \Delta(\boldsymbol{k}) \\
- \undertilde{\Delta}(\boldsymbol{k})  & -\undertilde{H}_{\mathrm{N}}(\boldsymbol{k})
\end{array}\right],
\end{align}
where $\undertilde{X}(\boldsymbol{k}, i\omega) \equiv X^\ast(-\boldsymbol{k}, i\omega)$ represents 
the particle-hole conjugation of $X(\boldsymbol{k}, i\omega)$.
Since zeros of the Pfaffian give the zeros of the excitation spectrum, the sign change of Pfaffian 
at a certain region in the Brillouin zone indicates the existence of quasiparticle states there.
A general expression of Pfaffian given in Eq.~(25) in Ref.~\citen{brydon:prb2018} is nonnegative 
for both the normal state and the superconducting states preserving time-reversal symmetry.
Unfortunately, it is not easy to obtain the analytical expression for the solutions of $P(\boldsymbol{k})<0$. 
We focus on a promising case of 
\begin{align}
\vec{\epsilon}_{\boldsymbol{k}} \cdot \vec{\eta}_{\boldsymbol{k}}=0,\label{eq:e-eta}
\end{align} 
under which the intraband pair potentials have nodes on the Fermi surface~\cite{brydon:prb2018}.
Indeed, numerical results showed that the BFSs exist around the nodes of the pair potentials.~\cite{brydon:prb2018}
The Pfaffian in such case is represented by, 
\begin{align}
P(\boldsymbol{k}) =&\left( \epsilon_{\boldsymbol{k},+}\, \epsilon_{\boldsymbol{k},-} 
+| \vec{\eta}_{\boldsymbol{k}}|^2 \right)^2 \nonumber\\
- 4 &\sum_{n>m>0}
|{\eta}_{\boldsymbol{k},n}|^2\, | {\eta}_{\boldsymbol{k},m}|^2
\sin^2\left( {\phi}_{\boldsymbol{k},n}-{\phi}_{\boldsymbol{k},m} \right),\\
\epsilon_{\boldsymbol{k},\pm} =& \xi_{\boldsymbol{k}} \pm |\vec{\epsilon}_{\boldsymbol{k}}|, \quad 
{\eta}_{\boldsymbol{k},n} = {\eta}^{\mathrm{Re}}_{\boldsymbol{k},n} e^{i {\phi}_{\boldsymbol{k}, n}},
\end{align}
where $\epsilon_{\boldsymbol{k},\pm}=0$ characterizes the Fermi surface in the normal state, 
${\eta}^{\mathrm{Re}}_{\boldsymbol{k},n}$ is a real function and ${\phi}_{\boldsymbol{k},n}$ 
represents a phase of the $n$ th component. 
The Pfaffian can be negative when the second term dominates the first one.
Although the first term is positive, the components of $\vec{\epsilon}_{\boldsymbol{k}}$ can decrease
the first term to be smaller than $| \vec{\eta}_{\boldsymbol{k}}|^2$. 
The second term remains finite only when the pair potential consists of more than one component 
and the relative phase of them 
${\phi}_{\boldsymbol{k},n}-{\phi}_{\boldsymbol{k},m}$ remains finite.
It is easy to confirm that such a superconducting state breaks time-reversal symmetry 
$\mathcal{T}\, \Delta(\boldsymbol{k})\, \mathcal{T}^{-1} \neq \Delta(\boldsymbol{k})$.

\begin{table*}[tb]
\caption{
Possible pairing states in cubic superconductors of $j=3/2$ fermions, which are decomposed into 
irreducible representations (irrep) of cubic symmetry ($O_h$).  Here, $1_{4 \times 4} \times U_T$ corresponds 
to a basis of singlet state, 
$\left(\mathcal{J}_1,\mathcal{J}_2,\mathcal{J}_3\right)$ to triplet states, 
$\left(\Gamma_{yz},\Gamma_{zx},\Gamma_{xy},\Gamma_{x^2-y^2},\Gamma_{3z^2-r^2} \right)$ to quintet states, 
and $\left(W_1,W_2,W_3,W_4,W_5,W_6,W_7\right)$ to septet states. 
In this paper, we consider even-parity $s$-wave order parameters ($\Delta(\bm{k})=\Delta$). 
The singlet and quintet pair can exist and form the potentials belonging to even-frequency symmetry because of $\Delta = -\Delta^T$.
On the other hand, the triplet and septet Cooper pairs cannot form the pair potentials because the permutation of two pseudospins in these states, $\Delta = \Delta^T$, contradicts the requirement from the Fermi-Dirac statistics of electrons.
Thus even-parity triplet pairs and even-parity septet pairs only appear as 
the pairing correlations belonging to odd-frequency symmetry class. }
\label{tab:pairing}
\begin{tabular}{clc}
\hline\hline
irrep  of $O_h$ & basis & pairing \\
\hline 
A$_{1g}$ & $1_{4\times 4} \times U_T$ & singlet \\
A$_{2g}$  & $W_7 = 1/\sqrt{3} (J_xJ_y J_z + J_z J_yJ_x) \times U_T$ & septet \\
E$_g$ & $\left(\Gamma_{x^2-y^2},\Gamma_{3z^2-r^2}\right) = \left[ 1/\sqrt{3} (J_x^2-J_y^2), 1/3 (2J_z^2-J_x^2-J_y^2) \right] \times U_T$ & quintet \\
T$_{1g}$  & $\left(\mathcal{J}_1,\mathcal{J}_2,\mathcal{J}_3\right)=\left( 2/\sqrt{5}J_x,2/\sqrt{5}J_y,2/\sqrt{5}J_z \right) \times U_T$ & triplet\\
           & $\left(W_1,W_2,W_3\right)=\left[ 2\sqrt{5}/3 \left(J_x^3-41/20 J_x \right), 2\sqrt{5}/3 \left(J_y^3-41/20 J_y \right),2\sqrt{5}/3 \left(J_z^3-41/20 J_z \right) \right] \times U_T$  & septet \\
T$_{2g}$  & $\left(\Gamma_{yz},\Gamma_{zx},\Gamma_{xy} \right)=\left[ 1/\sqrt{3}\left(J_yJ_z+J_z J_y\right), 1/\sqrt{3}\left(J_zJ_x+J_x J_z\right), 1/\sqrt{3}\left(J_xJ_y+J_y J_x\right) \right] \times U_T$ & quintet \\
            & $\left(W_4,W_5,W_6\right)=\left[ 1/\sqrt{3} \left\{ J_x,(J_y^2-J_z^2)\right\},  1/\sqrt{3} \left\{ J_y,(J_z^2-J_x^2)\right\},  1/\sqrt{3} \left\{ J_z,(J_x^2-J_y^2)\right\} \right] \times U_T$ & septet \\
\hline\hline
\end{tabular} 
\end{table*} 

\section{Odd-Frequency Cooper Pair }

 The Green's function for a superconducting state can be 
obtained by solving the Gor'kov equation
\begin{align}
&\left[\begin{array}{cc}
i\omega_n - H_{\mathrm{BdG}}(\boldsymbol{k}) 
\end{array}\right]
\left[\begin{array}{cc} G(\boldsymbol{k}, i\omega_n) & F(\boldsymbol{k}, i\omega_n) \\
 -\undertilde{F}(\boldsymbol{k}, i\omega_n) & -\undertilde{G}(\boldsymbol{k}, i\omega_n)
\end{array}\right] \nonumber\\
 &= 1_{8 \times 8}. \label{eq:gorkov}
\end{align}
The anomalous Green's function which represents the pairing correlation is calculated as
\begin{align}
F(\boldsymbol{k},i\omega_n) 
=& \left[ \undertilde{\Delta}+(i\omega_n 
+\undertilde{H}_{\mathrm{N}}) \Delta^{-1} (i\omega_n - H_{\mathrm{N}}) \right]^{-1},\label{eq:f-def}\\
%
%
=&\Delta\, \left[
\undertilde{\Delta} \Delta - \omega_n^2 - \undertilde{H}_{\mathrm{N}}\, \Delta^{-1}\, H_{\mathrm{N}}\, \Delta \right. \nonumber\\
&\left.
+i\omega_n \Delta^{-1} (\Delta \, \undertilde{H}_{\mathrm{N}} - H_{\mathrm{N}}\, \Delta)
\right]^{-1}.  
\end{align}
Thus the pairing correlation belonging to odd-frequency symmetry class exists 
when the relation
\begin{align}
\Delta \, \undertilde{H}_{\mathrm{N}} - H_{\mathrm{N}}\, \Delta \neq 0, \label{eq:odd_appear}
\end{align}
is satisfied.
In the present situation, 
we find $\undertilde{H}_{\mathrm{N}} = {H}^{\mathrm{T}}_{\mathrm{N}}$ in Eq.~(\ref{eq:hn_cubic}).
 Eq.~(\ref{eq:odd_appear}) is satisfied around the BFS because  
$\vec{\eta}_{\boldsymbol{k}} \neq 0$, $\vec{\epsilon}_{\boldsymbol{k}}\neq 0$ and Eq.~(\ref{eq:e-eta}) 
hold true simultaneously.~\cite{kim:prb2021} 
The condition in Eq.~(\ref{eq:odd_appear}) is realized in a multi-band/orbital 
superconductor often.\cite{BSchaffer:prb2013,asano:prb2015} 
It is not easy to have an analytical expression of the pairing correlation function in Eq.~(\ref{eq:f-def}).
Here, we assume that the temperature is near the transition temperature $T_c$
so that the relation $|\vec{\eta}_{\boldsymbol{k}}| \ll T_c$ holds true.
Within the first order of $|\vec{\eta}_{\boldsymbol{k}}|/T_c \ll 1$, the anomalous Green's function is calculated to be
\begin{align}
F(\boldsymbol{k}, i\omega_n)=& F_{\mathrm{even}}(\boldsymbol{k}, i\omega_n) + F_{\mathrm{odd}}(\boldsymbol{k}, i\omega_n),\\ 
F_{\mathrm{even}}(\boldsymbol{k}, i\omega_n) 
= &\frac{-1}{Z_{\mathrm{N}} }
\, \left( \omega_n^2 + \xi_{\boldsymbol{k}}^2 -|\vec{\epsilon}_{\boldsymbol{k}}|^2 \right) 
\, \vec{\eta}_{\boldsymbol{k}} \cdot \vec{\gamma} 
\, U_T, \\
F_{\mathrm{odd}}(\boldsymbol{k}, i\omega_n) 
= &\frac{-i \omega_n}{Z_{\mathrm{N}} }
 \, [\vec{\eta}_{\boldsymbol{k}}\cdot \vec{\gamma}, \vec{\epsilon}_{\boldsymbol{k}}\cdot \vec{\gamma}]_-
\, U_T, \label{eq:odd-f}\\
Z_{\mathrm{N}} =&(\omega_n^2 + \epsilon_{\boldsymbol{k},+}^2)(\omega_n^2 + \epsilon_{\boldsymbol{k},-}^2).
\end{align}
with $[a, b]_-$ being the commutator, where we have considered Eq.~(\ref{eq:e-eta}).  
The first term $F_{\mathrm{even}}$ belongs to pseudospin-quintet even-parity 
symmetry class and is linked to the pair potential. 
The second term $F_{\mathrm{odd}}$ belongs to odd-frequency symmetry class and satisfies
the relations,
\begin{align}
F_{\mathrm{odd}}(\boldsymbol{k}, -i\omega_n)=&-F_{\mathrm{odd}}(\boldsymbol{k},i\omega_n), \\
F_{\mathrm{odd}}(-\boldsymbol{k}, i\omega_n)=&F_{\mathrm{odd}}(\boldsymbol{k},i\omega_n),\\
F_{\mathrm{odd}}^{\mathrm{T}}(\boldsymbol{k}, i\omega_n)=&F_{\mathrm{odd}}(\boldsymbol{k},i\omega_n).
\end{align}
An odd-frequency Cooper pair exists regardless of whether time-reversal symmetry is preserved or broken.
Namely, the condition for the appearance of an odd-frequency pair in Eq.~(\ref{eq:odd_appear}) is looser 
than that of a quasiparticle on the BFS.
Therefore, a quasiparticle on the BFS could be relating to the special case of an odd-frequency pair.

\section{Odd-Frequency Pairs on Bogoliubov Fermi Surface}
The characters of an odd-frequency pair coexisting with a quasiparticle on the BFS can be analyzed by
assuming a practical pair potential breaking time-reversal symmetry~\cite{agterberg:prl2017}. 
We here focus on even-parity s-wave order parameters in cubic superconductors~\cite{brydon:prb2018}. 
We list the possible pairing states in a cubic structure in Table~\ref{tab:pairing}.
We chose the coordinate axes to be coincide with the cubic axes~\cite{PhysRev.102.1030}.
In $s$-wave symmetry, only the singlet and the quintet pair potentials 
satisfy the antisymmetric relation due to
the Fermi statistics of electrons $\Delta (\bm{k}) = -\Delta^T(-\bm{k})$.
The time-reversal symmetry breaking order parameters are allowed in pairing states described 
by two and three dimensional irreducible representations (irrep) in cubic symmetry $O_h$,
\begin{align}
{\rm E}_g: \Delta(\boldsymbol{k})&= \Delta (h_{3z^2-r^2}\Gamma_{3z^2-r^2}+ h_{x^2-y^2}\Gamma_{x^2-y^2}) \\
&= \Delta (h_{3z^2-r^2}\gamma^5+ h_{x^2-y^2}\gamma^4)\, U_T, \\
{\rm T}_{2g}: \Delta(\boldsymbol{k})&= \Delta (l_{yz}\Gamma_{yz}+l_{zx}\Gamma_{zx}+l_{xy}\Gamma_{xy}) \\
&= \Delta (l_{yz}\gamma^2+l_{zx}\gamma^3+l_{xy}\gamma^1)\, U_T,
\end{align}
where the E$_g$ and T$_{2g}$ pair potentials are 
characterized by the vectors $\bm{h} = (h_{3z^2-r^2},h_{x^2-y^2})$ and $\bm{l}=(l_{yz},l_{zx},l_{xy})$, respectively. 
The analysis of free energy suggests that the stable solutions breaking time-reversal symmetry 
are realized at $\bm{h}=(1,\pm i)$, $\bm{l}=(1,\pm i,0)$, $\bm{l}=(1,e^{2\pi i/3},e^{-2\pi i/3})$, and their 
equivalent solutions under cubic symmetry
~\cite{volvik:jetp1985,sigrist:rmp1991,brydon:prl2016}. 
In the following, we discuss the odd-frequency pairing correlations associated with those 
time-reversal symmetry breaking pair potentials.

\subsection{E$_g$ pairing order}
We consider the time-reversal symmetry breaking pair potential 
with $\bm{h}=(1,i \chi)$ for an E$_g$ symmetry. 
The pair potential is given by 
\begin{align}
\Delta(\boldsymbol{k})=&\Delta(\Gamma_{3z^2-r^2}+ i \chi \Gamma_{x^2-y^2})  
                            =\Delta(\gamma^5+ i \chi \gamma^4)\, U_T, \label{eq:eg_pair}
\end{align}
where $\chi =\pm 1$ indicates chirality of the pair potential. The corresponding odd-frequency pair is given by substituting Eq.~(\ref{eq:eg_pair}) to the odd-frequency component in Eq.~(\ref{eq:odd-f}),
\begin{align}
F_{\mathrm{odd}}(\boldsymbol{k}, i\omega_n) = &\frac{-i \omega_n \Delta}{Z_{\mathrm{N}} }
\left[ \gamma^5+ i \chi \gamma^4, \vec{\epsilon}_{\boldsymbol{k}} \cdot 
\vec{\gamma} \right]_- \, U_T. \label{eq:odd-f_eg}
\end{align}
An induced odd-frequency pairing correlation inherits the chirality of the pair potential. 
The resulting odd-frequency pairing correlation consists of triplet and septet states. 
This statement can be confirmed by decomposing Eq.~(\ref{eq:odd-f_eg}) into the irreducible spin matrices in Table~\ref{tab:pairing}\cite{boettcher:prb2017}. 
The details are relegated to Appendix\ref{app:decomp}. 
The odd-frequency pairing correlation is represented as
 \begin{align}
 &F_{\mathrm{odd}}(\boldsymbol{k}, i\omega_n) =  \frac{-2i \omega_n \Delta}{Z_{\mathrm{N}} } \notag \\
 &\times \Bigg\{ \frac{\chi}{\sqrt{5}} \Big[ \epsilon_{\bm{k},2} \mathcal{J}_1 +\epsilon_{\bm{k},3} \mathcal{J}_2 -2 \epsilon_{\bm{k},1} \mathcal{J}_3 
 + i \chi \sqrt{3} (\epsilon_{\bm{k},3} \mathcal{J}_2 - \epsilon_{\bm{k},2} \mathcal{J}_1)\Big] \notag  \\
  &-\frac{\chi}{2\sqrt{5}} \Big[ \epsilon_{\bm{k},2} W_1 +\epsilon_{\bm{k},3} W_2 -2 \epsilon_{\bm{k},1} W_3 
+ i \chi \sqrt{3} (\epsilon_{\bm{k},3} W_2 - \epsilon_{\bm{k},2} W_1)\Big] \notag  \\
 &
 +\frac{\chi}{2} \Big[ \sqrt{3}(\epsilon_{\bm{k},2} W_4 - \epsilon_{\bm{k},3} W_5) 
 + i \chi (\epsilon_{\bm{k},2} W_4 +\epsilon_{\bm{k},3} W_5 -2\epsilon_{\bm{k},1} W_6) \Big] \notag  \\
 &+\chi (\epsilon_{\bm{k},5}+i \chi \epsilon_{\bm{k},4}) W_7 \Bigg\}. \label{eq:odd-f_eg_decomp}
 \end{align}
 The first line in Eq.~(\ref{eq:odd-f_eg_decomp}) describes chiral odd-frequency triplet pairs.
All the remaining terms describe chiral odd-frequency septet pairs. 
The emergence of the odd-frequency septet pairs features superconductivity of $j=3/2$ fermions. 
It would be worth mentioning that the last term is proportional to an 
octopolar magnetic order parameter coming from the time-reversal-odd gap product~\cite{brydon:prb2018}.
Therefore, chiral odd-frequency septet pairs are related to a nonunitary pairing state.

\begin{figure}[t]
\begin{center}
  \includegraphics[width=7.0cm]{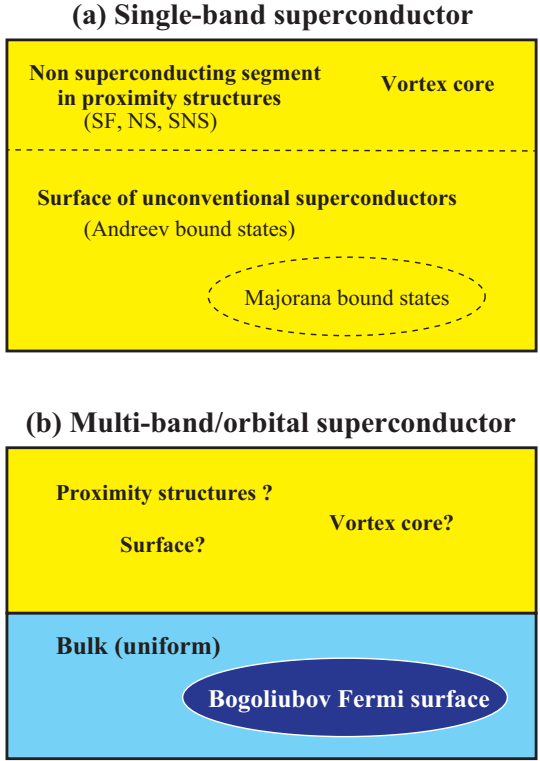}
\caption{(Color online) Places where an odd-frequency Cooper pair exists (a) in a single-band superconductor
and (b) in a two-band/orbital superconductor.
(a) Spatially uniform odd-frequency pairing correlations are absent in a single-band superconductor.
Therefore, an odd-frequency pair exists as the subdominant correlations which spatially localize 
at a junction interface, a surface and a vortex core. 
The odd-frequency pairing correlations in such places represent the local deformation 
of the superconducting condensate.
(b) In two-band/orbital superconductors, an odd-frequency pair exists as a spatially uniform subdominant pairing correlation in the bulk as well as a local deformation of the condensate.
A quasiparticle on the BFS appears due to extra degrees of freedom in the electronic structure in the absence of time-reversal symmetry~\cite{agterberg:prl2017}.
The physical phenomena caused by an odd-frequency pair in the bulk have been an open issue.
In this paper, we show that a quasiparticle on the BFS is related strongly to an odd-frequency 
Cooper pair characterized by chirality.
 }
\label{fig1}
\end{center}
\end{figure}

\subsection{T$_{2g}$ pairing order (chiral state)}
In the case of T$_{2g}$ symmetry, 
time-reversal symmetry breaking states are classified into two types: 
chiral state $\bm{l}=(1,i \chi,0)$ ($\chi = \pm 1$) and 
cyclic state $\bm{l}=(1,\omega,\omega^{-1})$ ($\omega = e^{i2\pi/3}$). 
They are sixfold and eightfold degenerate, respectively.~\cite{brydon:prb2018}  We first examine odd-frequency pairs associated with the chiral state, whose pair potential is given by
\begin{align}
\Delta(\boldsymbol{k})=&\Delta(\Gamma_{zx}+ i \chi \Gamma_{yz})  
                            =\Delta(\gamma^3+ i \chi \gamma^2)\, U_T. 
\end{align}
We find that the resulting odd-frequency pairing correlation in such a chiral superconductor
\begin{align}
F_{\mathrm{odd}}(\boldsymbol{k}, i\omega_n) = &\frac{-i \omega_n \Delta}{Z_{\mathrm{N}} }
\left[ \gamma^3 +i \chi \gamma^2, \vec{\epsilon}_{\boldsymbol{k}} \cdot 
\vec{\gamma} \right]_- \, U_T,
\end{align}
 also inherits the chirality of the pair potential. 
 The odd-frequency pairs can be decomposed into triplet and septet pairing states as
 \begin{align}
F_{\mathrm{odd}}&(\boldsymbol{k}, i\omega_n) =  \frac{-2i \omega_n \Delta}{Z_{\mathrm{N}} }\notag \\
&\times \Bigg\{\frac{\chi}{\sqrt{5}}\Big[ -\epsilon_{\bm{k},1}\mathcal{J}_2-(\epsilon_{\bm{k},4}+\sqrt{3}\epsilon_{\bm{k},5})\mathcal{J}_1 \notag \\
&\quad - i \chi (\epsilon_{\bm{k},1}\mathcal{J}_1-(\epsilon_{\bm{k},4}-\sqrt{3}\epsilon_{\bm{k},5})\mathcal{J}_2)\Big] \notag \\
&+\frac{\chi}{2\sqrt{5}}\Big[ -4 \epsilon_{\bm{k},1}W_2+(\epsilon_{\bm{k},4}+\sqrt{3}\epsilon_{\bm{k},5})W_1 \notag \\
&\quad- i \chi (4 \epsilon_{\bm{k},1}W_1+(\epsilon_{\bm{k},4}-\sqrt{3}\epsilon_{\bm{k},5})W_2)\Big] \notag \\
&+\frac{\chi}{2}\Big[ (\epsilon_{\bm{k},5}-\sqrt{3}\epsilon_{\bm{k},4})W_4 \notag \\
&\quad- i \chi (\epsilon_{\bm{k},5}+\sqrt{3}\epsilon_{\bm{k},4}) W_5\Big]\notag  \\
&+\frac{\chi}{\sqrt{5}} (\epsilon_{\bm{k},3}+i \chi \epsilon_{\bm{k},2}) (\mathcal{J}_3+2W_3)\Bigg\}. \label{eq:odd-f_t2gchiral_decomp}
\end{align}
The last term in Eq.~(\ref{eq:odd-f_t2gchiral_decomp}) is proportion to $4 J_z^3-7J_z$, which again describes the time-reversal-odd gap product.

\subsection{T$_{2g}$ pairing order (cyclic state)}
Next, we consider the cyclic pair potential,
\begin{align}
 \Delta(\bm{k}) &= \Delta (\Gamma_{yz}+\omega \Gamma_{zx}+\omega^{-1} \Gamma_{xy}) \\
&= \Delta (\gamma^2+\omega \gamma^3+\omega^{-1} \gamma^1)\, U_T,
\end{align}
which yields the odd-frequency pairing correlation,
\begin{align}
F_{\mathrm{odd}}(\boldsymbol{k}, i\omega_n) = &\frac{-i \omega_n \Delta}{Z_{\mathrm{N}} }
\left[ \gamma^2+\omega \gamma^3+\omega^{-1} \gamma^1, \vec{\epsilon}_{\boldsymbol{k}} \cdot 
\vec{\gamma} \right]_- \, U_T.
\end{align}
The pairing correlation function involves the phase factors $\bm{l}=(1,\omega,\omega^{-1})$. 
The results of the decomposition are given by
\begin{align}
 F_{\mathrm{odd}}&(\boldsymbol{k}, i\omega_n) =  \frac{-2i \omega_n \Delta}{Z_{\mathrm{N}} } \notag \\ 
 &\times \Bigg\{ \frac{i}{\sqrt{5}}\Big[(\epsilon_{\bm{k},4}+\sqrt{3}\epsilon_{\bm{k},5})\mathcal{J}_1+\omega (\epsilon_{\bm{k},4}-\sqrt{3}\epsilon_{\bm{k},5})\mathcal{J}_2 \notag \\
 &\qquad  \quad  - 2 \omega^{-1} \epsilon_{\bm{k},4} \mathcal{J}_3 \Big]  \notag \\
 &-\frac{i}{2\sqrt{5}}\Big[(\epsilon_{\bm{k},4}+\sqrt{3}\epsilon_{\bm{k},5})W_1+\omega (\epsilon_{\bm{k},4}-\sqrt{3}\epsilon_{\bm{k},5})W_2 \notag \\
&\qquad  \quad - 2 \omega^{-1} \epsilon_{\bm{k},4} W_3 \Big]  \notag \\
 &-\frac{i}{2}\Big[(\epsilon_{\bm{k},5}-\sqrt{3}\epsilon_{\bm{k},4})W_4+\omega (\epsilon_{\bm{k},5}+\sqrt{3}\epsilon_{\bm{k},4})W_5 \notag \\
&\qquad  \quad  - 2 \omega^{-1} \epsilon_{\bm{k},5} W_6 \Big]  \notag \\
 +&\frac{i}{\sqrt{5}} \Big[ \epsilon_{\bm{k},1} (\mathcal{J}_2+2W_2)-\epsilon_{\bm{k},3} (\mathcal{J}_3+2W_3)\notag \\
 &+\omega (\epsilon_{\bm{k},2} (\mathcal{J}_3+2W_3)-\epsilon_{\bm{k},1} (\mathcal{J}_1+2W_1))\notag \\
 &+\omega^{-1}(\epsilon_{\bm{k},3} (\mathcal{J}_1+2W_1)-\epsilon_{\bm{k},2} (\mathcal{J}_2+2W_2))\Big]
 \Bigg\}, \label{eq:odd-f_t2gcyclic_decomp}
\end{align}
which consists of both the triplet and septet components. 
The last three terms indicate the terms proportional to $4 J_i^3 -7 J_i$ inherent to the time-reversal odd product.

\section{Discussion}
Historically, Berezinskii~\cite{berezinskii:jetplett1974} proposed the pair potential 
belonging to odd-frequency symmetry class (odd-frequency superconductivity/superfluidity). 
Today, however, we know that spatially uniform single-band odd-frequency superconductivity is 
impossible~\cite{fominov:prb2015}. 
In Fig.~\ref{fig1}, we compare characters of an odd-frequency Cooper pair 
in a single-band superconductor and those in a multi-band/orbital superconductor.
In the case of single-band, an odd-frequency Cooper pair appears as a 
 subdominant pairing correlation which localizes various places such as a vortex core, a surface and
a junction interface to another material as shown in Fig.~\ref{fig1}(a).
The odd-frequency pairing correlations in these cases describe 
the local deformation of the superconducting condensate~\cite{asano:prb2014}.
A spin-triplet $s$-wave Cooper pair generated by the exchange potential in a ferromagnet
~\cite{bergeret:prl2001} is the most well-known example of an odd-frequency pair.
The exchange potential polarizes and flips spin of an electron, which causes the symmetry conversion 
between spin-singlet and spin-triplet.
At a surface of an unconventional superconductor, Andreev bound states appear 
due to the sign change of the pair potential. 
The surface breaks inversion symmetry locally and generates an odd (even)-parity Cooper pair 
from an even (odd)-parity pair in the bulk~\cite{tanaka:prl2007,asano:prb2014}. 
Majorana fermion which is a quasiparticle at a specialized Andreev bound state also accompanies 
an odd-frequency pair~\cite{asano:prb2013}. 
As far as we know, an odd-frequency pair exhibits paramagnetic response to an external magnetic field~\cite{tanaka:prb2005,asano:prl2011,asano:prb2015}.
Thus the appearance of an odd-frequency pair drastically modifies local magnetic properties 
such as the surface impedance of a junction~\cite{asano:prl2011} and the magnetic susceptibility 
of a small unconventional superconductor~\cite{suzuki:prb2014}. 

In contrast to single-band superconductors, the odd-frequency pairing correlation 
exists as a part of the uniform ground state~\cite{BSchaffer:prb2013} as shown in Fig.~\ref{fig1}(b).  
With the following exceptions, the nature of an odd-frequency pair in the bulk is not yet well understood.
An odd-frequency pair suppresses $T_c$~\cite{asano:prb2015} because it decreases the pair density and the 
condensation energy. Odd-frequency pairs induced in the bulk are known to stabilize the 
Josephson $\pi$-state.~\cite{sasaki:prb2020,kim:prb2021}. 
At present, we have never known any properties of a chiral odd-frequency pair discussed in Sec.~4. 
Investigation in such a direction would make it possible to understand physics of a quasiparticle on BFS.
Finally, we found in a very recent paper that the authors mention only
the relation between an odd-frequency pair and a quasiparticle on the BFS.~\cite{dutta} 
The characters of an odd-frequency pair were not discussed at all.


\section{Conclusion}
We discussed a relationship between a quasiparticle on the Bogoliubov Fermi surface (BFS)
 and an odd-frequency Cooper pair in a two-band/orbital superconductor which breaks 
time-reversal symmetry. 
By solving the Gor'kov equation analytically, we find that a quasiparticle on the 
BFS accompanies an odd-frequency Cooper pair characterized by chirality. 
We also analyze symmetry of odd-frequency pairs which are induced in a cubic superconductors.

\begin{acknowledgment}
The authors are grateful to S.~Ikegaya for useful discussion.
This work was supported by JSPS KAKENHI (No. JP20H01857), 
JSPS Core-to-Core Program (No. JPJSCCA20170002), 
and JSPS and Russian Foundation for Basic Research under Japan-Russia Research Cooperative Program
Grant No. 19-52-50026. 
S. K. was supported by JSPS KAKENHI Grant No. JP19K14612 and by the CREST project
(JPMJCR16F2, JPMJCR19T2) from Japan Science and Technology Agency (JST).
\end{acknowledgment}

\appendix

\section{Decomposition Formulae for Odd-Frequency Pairs}
\label{app:decomp}
Here we summarize the decomposition of odd-frequency pairing correlation function 
into triplet and septet components in $j=3/2$ superconductors with cubic symmetry. 
We used the following relations
\begin{align}
 &\gamma^{1}\gamma^2\;U_T = -\frac{i}{\sqrt{5}}(\mathcal{J}_2+2W_2), \ \  \gamma^{1}\gamma^3\;U_T = \frac{i}{\sqrt{5}}(\mathcal{J}_1+2W_1), \notag \\
  &\gamma^{1}\gamma^4\;U_T = -\frac{i}{\sqrt{5}}(\mathcal{J}_3-2W_3), \ \  \gamma^{1}\gamma^5\;U_T =i W_6, \notag \\
  &\gamma^{2}\gamma^3\;U_T = -\frac{i}{\sqrt{5}}(\mathcal{J}_3+2W_3), \notag \\  
  &\gamma^{2}\gamma^4\;U_T =\frac{i}{\sqrt{5}}\mathcal{J}_1-\frac{i}{2\sqrt{5}}W_1+i\frac{\sqrt{3}}{2}W_4, \notag \\
  &\gamma^{2}\gamma^5\;U_T = i\sqrt{\frac{3}{5}} \mathcal{J}_1-\frac{i}{2} \sqrt{\frac{3}{5}} W_1-\frac{i}{2} W_4, \notag \\
  &\gamma^{3}\gamma^4\;U_T =\frac{i}{\sqrt{5}} \mathcal{J}_2 -\frac{i}{2\sqrt{5}} W_2-i\frac{\sqrt{3}}{2}W_5, \notag \\
  &\gamma^{3}\gamma^5\;U_T = -i\sqrt{\frac{3}{5}} \mathcal{J}_2+\frac{i}{2} \sqrt{\frac{3}{5}} W_2-\frac{i}{2} W_5, \notag \\
  &\gamma^{4}\gamma^5\;U_T = -i W_7,
\end{align}
in Eqs.~(\ref{eq:odd-f_eg_decomp}), (\ref{eq:odd-f_t2gchiral_decomp}), and (\ref{eq:odd-f_t2gcyclic_decomp}).


\end{document}